\documentclass[amsmath,amssymb,a4paper]{article}
\usepackage[linktocpage=true]{hyperref}
\usepackage{graphicx}
\usepackage{booktabs}
\usepackage{epsfig}
\usepackage{color}
\usepackage{epstopdf}
\usepackage{authblk}
\newcommand{\E}{\mathcal{E}}
\newcommand{\T}{\mathcal{T}}
\newcommand{\red}[1]{#1} 

\begin{document}
\title{Correlations between charge and energy current 
in  ac-driven coherent conductors} 
\author[1]{Francesca Battista}

\author[2]{Federica Haupt}
\author[3]{Janine Splettstoesser}
\affil[1]{Departamento de F\'isica, FCEyN, Universidad de Buenos Aires and IFIBA,
Pabell\'on I, Ciudad Universitaria, 1428 CABA Argentina}
\affil[2]{JARA Institute for Quantum Information, RWTH Aachen University, D-52056 Aachen, Germany}
\affil[3]{Department of Microtechnology and Nanoscience (MC2), Chalmers University of Technology, SE-41298 G\"oteborg, Sweden}

\date{}
  
 \maketitle
\begin{abstract}
We study transport in coherent conductors driven by a time-periodic bias voltage. We present results of the charge and energy noise and  complement them by a  study  of the mixed noise, namely the zero-frequency correlator between charge and energy current.  The mixed noise presents interference contributions and transport contributions, showing features different from those of charge and energy noise. The mixed noise can be accessed by measuring the correlator between the fluctuations of the power provided to the system and the charge current.
\end{abstract}
\section{Introduction} 
In the last years, big effort has been put in the theoretical study and in the experimental investigation of charge transport in mesoscopic structures. Single particle sources, allowing for the  control in time and space of the flow of a small number of electrons or holes, have been realized with time-periodically  driven mesoscopic systems~\cite{Pekola13rev}. These setups are usually based on the emission of particles from a confined region.

Levitov and coworkers~\cite{Keeling06} proposed an alternative way to achieve single-particle emission without using quantum confinements. Special  Lorentzian-shaped voltage pulses applied to a conductor give  rise only to single-particle excitations from a Fermi sea, free from electron-hole pairs excitations. Such  Lorentzian pulse $V(t)$ satisfies the condition $\int_0^{\mathcal{T}}dt eV(t)/h=l$ where $l$ is an integer number. Experimentally, Lorentzian pulses carrying an integer number of particles can approximately  be realized by superposing several harmonic driving potentials. In this case a reduction of the charge current noise was measured in Refs.~\cite{Gabelli13,Dubois13nat}. 

Charge current noise in ac-driven systems has  been widely studied. However the charge carriers, electrons and holes, transport energy as well. It is thus important for future applications to complement the existing literature (see e.g.~\cite{Gabelli13,chlit,Vanevic,Reydellet03}) 
with an accurate study of energy noise in these systems~\cite{Battista14}.  Recently, energy noise in normal conducting~\cite{Crepieux14} and superconducting systems~\cite{Virtanen14} has been studied in the stationary regime; also fluctuation relations in  systems with a time-dependent driving have been considered~\cite{Moskalets14,Giazotto14}. 
In this paper, we model the setups of the latest experiments~\cite{Gabelli13,Dubois13nat} with a two-terminal conductor with a central scatterer. 
After reporting results~\cite{Battista14} on the correlator of charge current fluctuations (charge noise) and the correlator of energy current fluctuations (energy noise)  we calculate and discuss the behavior of the correlator between the charge and the energy current fluctuations (mixed noise) \cite{Crepieux14}. 

\section{Formalism and model}\label{sec:model}
Our considerations are based on the scattering theory for  photon-assisted transport~\cite{Pretre96}.
We consider a coherent mesoscopic conductor connected to metallic contacts (reservoirs) by ballistic leads. The reservoirs are subjected to time-periodic voltages $V_{\alpha}(t)$.
The charge  current operator in contact $\alpha$ is~\cite{Buttiker92}  
\begin{equation}\label{eq:current}
\hat{I}^{\mathcal{C}}_{\alpha}(t)=\frac{e}{h} \int^{\infty}_{-\infty}\int^{\infty}_{-\infty} dE dE' e^{i(E-E')t/\hbar}\hat{i}_{\alpha}(E,E'),
\end{equation} 
while the corresponding energy current operator is~\cite{Ludovico13}
\begin{equation}\label{eq:Ecurrent}
\hat{I}^{\mathcal{E}}_{\alpha}(t)
=\frac{1}{h} \int^{\infty}_{-\infty}\int^{\infty}_{-\infty} dE dE' \frac{(E+E')}{2}e^{i(E-E')t/\hbar}\hat{i}_{\alpha}(E,E').
\end{equation}
Here,  $e<0$ is the electron charge and $\hat{i}_{\alpha}(E,E')=\hat{\bf b}^{\dagger}_{\alpha}(E)\hat{\bf b}_{\alpha}(E')-\hat{\bf a}^{\dagger}_{\alpha}(E)\hat{\bf a}_{\alpha}(E')$. The vectors $\hat{\bf a}_{\alpha}$ and  $\hat{\bf b}_{\alpha}$ have the operators $\hat{a}_{\alpha n}$ and $\hat{b}_{\alpha n}$ as their components, annihilating an electron in channel $n$ in lead $\alpha$ moving towards or away from the scatterer, respectively. 
Of our interest are the zero-frequency auto- and cross-correlators of charge current (charge noise), $S_{\alpha\beta}$, of energy current (energy noise), $S_{\alpha \beta}^{\E}$, as well as the correlators between charge and energy current (mixed noise), $S_{\alpha \beta}^{X}$, which are expected to be nonzero because electrons and holes carry both charge and energy. In the case of time periodic driving considered here, we can write in energy space~\cite{Rychkov05}
\begin{equation}\label{eq:noise-gen} 
S_{\alpha\beta}^{xy}=h  \int^{\infty}_{-\infty} \int^{\infty}_{-\infty} dE  dE' \langle \Delta \hat{I}^{x}_{\alpha}(E)\Delta \hat{I}^{y}_{\beta}(E') \rangle,
\end{equation}
with $\Delta\hat{A}=\hat{A}-\langle \hat{A}\rangle$, and the charge current density, $\hat{I}^{\mathcal{C}}_{\alpha}(E)=e/h\cdot \hat{i}_{\alpha}(E,E)$, and energy current  density, $\hat{I}_{\alpha}^{\E}(E)=E/h\cdot \hat{ i}_{\alpha}(E,E)$. This equation yields the charge noise for $x=y=\mathcal{C}$, the energy noise for $x=y=\mathcal{E}$, as well as the mixed noise for \red{$x=\mathcal{C},y=\mathcal{E}$}. 
The quantum statistical averages, $\langle\hat{A}\rangle$, in Eq.~(\ref{eq:noise-gen}) can be evaluated substituting $\hat{\bf b}_{\alpha}(E)=\sum_{\beta}{\bf s}_{\alpha \beta}(E)\hat{\bf a}_{\beta}(E)$, where ${\bf s}_{\alpha \beta}(E)$ is the scattering matrix relating $\hat{\bf a}_{\alpha}$ to $\hat{\bf b}_{\alpha}$, and for leads with $N_{\alpha}$ and $N_{\beta}$ channels has dimensions $N_{\alpha}\times N_{\beta}$.
Interacting with $V_{\alpha}(t)$, the electrons in the reservoir $\alpha$ can absorb $k$ energy quanta $\hbar\Omega$ with probability amplitude $ c_{\alpha k}=\int_{0}^{\T} \frac{dt}{\T}\, e^{-i\frac{e}{\hbar}\int_{0}^{t} dt' [V_{\alpha}(t')-\overline{V}_{\alpha}]}\,e^{ik\Omega t}$,
 where $\Omega=2\pi/\T$ is the frequency of the driving and $\overline{V}_{\alpha}$ is the dc component of $V_{\alpha}(t)$.
A state with energy $E$ in the leads corresponds to a superposition of  reservoir states with energy $E_{-k}=E-k\hbar\Omega$~\cite{Pretre96}.
The statistics of the operators  $\hat{a}_{\alpha n}(E)$ and $\hat{a}_{\alpha n}^{\dag}(E)$ is thus  
\begin{equation}
\langle \hat{a}_{\alpha n}^{\dag}(E) \hat{a}_{\beta m}(E')  \rangle=\delta_{\alpha\beta}\delta_{mn}\sum_{k,\ell=-\infty}^{+\infty}\!\!\! c^{*}_{\alpha k}c_{\beta k+\ell}f_{\alpha}(E_{-k})\delta(E_{\ell}\!-\!E'), 
\end{equation} 
 with the Fermi function $f_\alpha(E)=\left[1+\exp\{(E-eV_\alpha)/k_{\rm B}T\}\right]^{-1}$, the temperature $T$ and the Boltzmann constant $k_\mathrm{B}$.
 As a consequence of the unitarity of the
scattering matrix we have that $S^{xy}_{LL}=S^{xy}_{RR}=-S^{xy}_{LR}=-S^{xy}_{RL}$. For simplicity, from now on we will  always refer to the auto-correlators in the right reservoir $S\equiv S_\mathrm{RR}$, $S^{\mathcal{E}}=S^{\mathcal{E}}_\mathrm{RR}$ and $S^{X}\equiv S^X_\mathrm{RR}$.
 
Inspired by the experiments in Refs.~\cite{Gabelli13,Dubois13nat}, we consider a two-terminal conductor and  we assume an energy-independent transmission  $D$ of the scatterer~\cite{Gabelli13}. In order to simplify the notation, we focus on a spinless single-channel conductor. The left contact is subject to a periodic time-dependent potential $V_{\rm L}(t)=V_{\rm ac}(t)+\bar{V}$, where $\bar{V}$ is a dc voltage offset and $V_{\rm ac} (t)$ is a pure ac component. The right contact is grounded $V_{\rm R}(t)=0$, and  energies are taken with respect to the electrochemical potential $\mu=0$ of this reservoir.\footnote{Note that with this choice the energy current in the right reservoir coincides with the heat current.} The time-dependent driving excites the Fermi sea of the left reservoir.
Because of Pauli exclusion principle, only occupied states above $\mu$ participate to transport and thus we define the excitations above $\mu$ as electron-like. Empty states below $\mu$ contribute to transport as well. We refer to the excitations below $\mu$ as hole-like excitations. 
Correspondingly, we write the energy resolved current  operators as the sum of two contributions, one carried by electron-like and the other by hole-like excitations,  $\hat{I}^\mathcal{C}_\alpha(E)=\sum_{i=\rm e, h}\hat{I}_\alpha^{\mathcal{C}(i)}(E)$, and $\hat{I}_\alpha^{\mathcal{E}}(E)=\sum_{i=\rm e, h}\hat{I}_\alpha^{\mathcal{E}(i)}(E)$~\cite{Rychkov05}.

\section{Results}\label{sec:results}
We want to investigate the correlations between charge and energy currents carried by the electron and hole excitations participating to transport. It is instructive to decompose the charge, energy and mixed noise in terms of contributions that account for the correlations between excitations of the same or of different kind~\cite{Battista14} 
\begin{equation}\nonumber
S^{xy (ij)}=h \int^{\infty}_{-\infty}\int^{\infty}_{-\infty} dE dE' \langle \Delta \hat{I}^{x(i)}(E)\Delta \hat{I}^{y(j)}(E') \rangle.
\end{equation}
These correlators show two contributions, $S^{xy (ij)}=S^{xy (ij)}_{\rm tr}+S^{xy (ij)}_{\rm int}$, see Ref.~\cite{Battista14}, where
\begin{eqnarray} \label{eq:Sxytr}
S^{xy(ij)}_{\mathrm{tr}}\!\!\!\!&=&\!\!\!\!\! \frac{D(1-D)}{h}\!\! \int \!\! dE\, \tilde{x}\tilde{y} \! \! \! \sum_{\ell=-\infty}^{\infty}\! |c_\ell|^2 \left\{ f_\mathrm{L}(E_{-\ell})\left[1\!-\! f_\mathrm{R}(E)\right]\!+\! f_\mathrm{R}(E)\big[1\! -\! f_\mathrm{L}(E_{-\ell})\big]\right\} \theta_{i}(E)\theta_{j}(E),\\  \label{eq:Sxyint}
S^{xy(ij)}_{\mathrm{int}}\!\!\!\! &=& \!\!\!\! \frac{D^2}{h}\!\!\sum_{\alpha=\rm L,R}\sum_{k,\ell,q=-\infty}^{\infty}\!\!\! c_{\alpha\ell}^{*}c_{\alpha(\ell+q)}c_{\alpha(k+q)}^{*}c_{\alpha k}\int \!\! dE\,\tilde{x}\tilde{y_q} f_\alpha(E_{-\ell})\left[1-f_\alpha(E_{-k})\right]\theta_{i}(E)\theta_{j}(E_q),
\end{eqnarray}
with distinct physical origins. The expressions for the charge, energy, and mixed noise, are obtained by replacing $\tilde{x}=e$ ($\tilde{x}=E$) when $x=\mathcal{C}$ ($x=\mathcal{E}$) and $\tilde{y}=e$ and $\tilde{y_q}=e$  ($\tilde{y}=E$ and $\tilde{y_q}=E_q$) when $y=\mathcal{C}$ ($y=\mathcal{E}$). We introduced $\theta_{\rm e/h}(E)=\theta(\pm E)$ with the Heaviside step function $\theta(E)$.
The first term, Eq.(\ref{eq:Sxytr}), depends on the Fermi distribution of the two reservoirs and it is related to correlations due to the exchange of particles between the two contacts. We call it transport part of the noise~\cite{Battista14}, and it is nonzero only if one considers correlations between the same type of excitations, i.e. $S^{ xy\rm (eh)}=S^{xy \rm (he)}=0$. This indicates that  electrons and holes contribute independently to the transport terms. The scatterer  randomly transmits and reflects charge and energy carriers and this is reflected in the factor $D(1-D)$. 
The second term, Eq.(\ref{eq:Sxyint}) originates ultimately from correlations due to the exchange of particles between states with different energies in the same reservoir. Without periodic driving, only elastic exchange processes (i.e. thermal fluctuations) contribute to the noise, since $c_{{\rm L} k}=\delta_{k0}$ if  $V_{\rm ac}=0$.  In the presence of a periodic driving, correlations between states with different energies $E_\ell$, $E_k$ do in general contribute to the charge, energy and mixed noise. However, how large  this contribution is depends on the interference between different ``paths" 
in energy space, that electrons can take when being promoted from one energy state to another by interacting with the ac-field. 
Therefore  we refer to $S_{\rm int}^{xy(ij)}$ as the interference part of the noise~\cite{Battista14}. 
 
 In the following we summarize the main features of the interference and the transport part of the charge and energy noise studied in Ref.~\cite{Battista14}, and then focus on the mixed noise.   In all cases, we make examples taking the drivings used in Ref.~\cite{Gabelli13}:  a simple harmonic driving $V^{\rm h}_{\rm L}(t)=\bar{V}+V_{0}\cos(\Omega t)$ and a bi-harmonic driving $V^{\rm bh}_{\rm L}(t)=\bar{V}+V_{0}\cos(\Omega t)+\frac{V_{0}}2\cos(2\Omega t)$.
\section{Charge noise}\label{sec:chargen}
We first analyse the interference part of the charge noise,  $S_{\rm int}=\sum_{ij}S_{\rm int }^{(ij)}$.
It can be shown that this simply corresponds  to  thermal fluctuations $S_{\rm int}=2\frac{e^{2}}{h}D^{2}k_{\rm B}T$, indicating that, even in the presence of an ac-field, inelastic exchange processes in one reservoir interfere destructively when they all contribute to a certain observable with the same weight ($e^2$, for the case of the charge noise)~\cite{Battista14}. 
In  Fig.~\ref{fig}a) we plot the interference parts of the different contributions $S^{(ij)}$ as a function of the dc component of the bias at zero temperature. We find $S^{(\rm ee)}_{\rm int}=S^{(\rm hh)}_{\rm int}=-S^{(\rm eh)}_{\rm int}=-S^{(\rm he)}_{\rm int}$ giving indeed $S_{\rm int}(k_{\rm B}T=0)=0$. 

We now consider the transport contribution to the charge noise, $S_{\rm tr}=\sum_{i}S_{\rm tr}^{(ii)}$, given by
\begin{equation} 
S_{\mathrm{tr}} = \frac{e^2D(1-D)}{h} \sum_{\ell=-\infty}^{+\infty}|c_{\rm L \ell}|^{2}(\ell\hbar\Omega+e\bar{V})\coth\left(\frac{\ell\hbar\Omega+e\bar{V}}{2k_{\rm B}T}\right). \label{eq:StrT}
\end{equation}
At zero temperature this is the total charge noise $S_{\mathrm{tr}}(k_{\rm B}T=0)=S(k_{\rm B}T=0)$ and it  can be written as the sum of a contribution due to  electron-like excitations and a contribution due to hole-like excitations~\cite{Reydellet03},   
\begin{equation}\label{eq:SiiT0}
S(k_{\rm B}T=0)=\sum_{i=\rm e,h}\frac{e^{2}\Omega}{2\pi}D(1-D)N_{i}\ .
\end{equation}
Here, $N_{i}=\sigma_{i}\sum_{\ell=-\infty}^{+\infty}|c_{\rm L \ell}|^{2}(\ell+e\bar{V}/\hbar\Omega)\theta_{i}(\ell \hbar\Omega+e\bar{V})$ is the number of electrons ($i=\rm e$) or holes ($i=\rm h$) that arrive on the scatterer during one period, with $\sigma_{\rm e/h}=\pm$.  
Gabelli and Reulet~\cite{Gabelli13} observed a reduction of the shot noise in a tunnel junction driven by a bi-harmonic signal with respect to the shot noise due to a simple harmonic driving. This  is a consequence of the inhibition of  electron-hole pair creation~\cite{Vanevic}. 
In Fig.~\ref{fig}b) we plot the transport contributions to the charge correlator  $S_{\rm tr}^{(ii)}$ at zero temperature. It indeed shows a minimum in the noise when a biharmonic signal is applied, which corresponds to a reduction of  $S^{\rm(ee)}_{\rm tr}$ and $S^{\rm(hh)}_{\rm tr}$ with respect to the case of  harmonic driving. 
\section{Energy noise}\label{sec:energyn} 
\begin{figure}[h!]
\begin{center}
\includegraphics[width=0.95\textwidth]{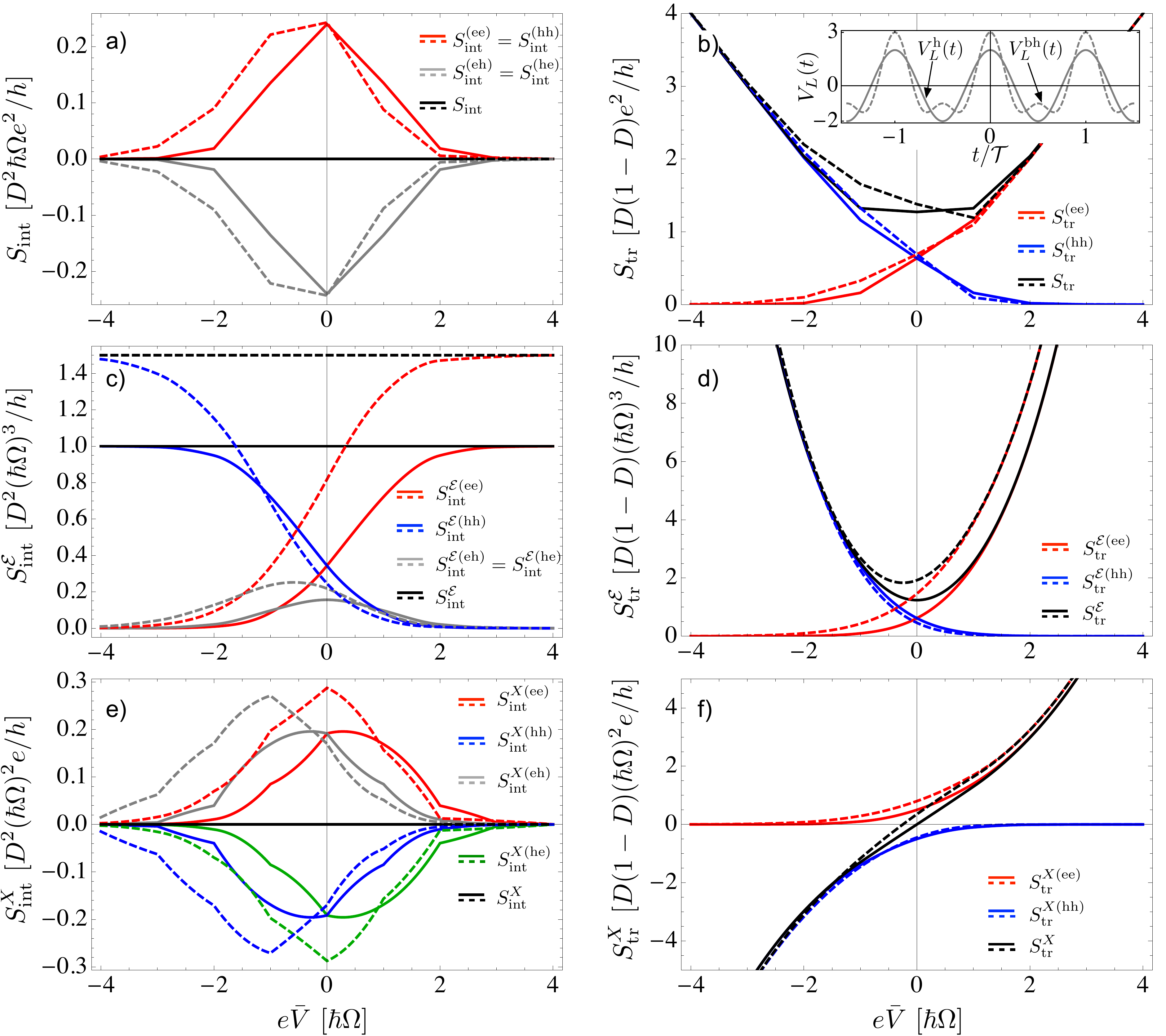}
\end{center}
\caption{Plots of the interference and transport  contributions to the charge ($S_{\rm int}^{(ij)},S_{\rm tr}^{ (ij)}$), energy ($S_{\rm int}^{\E (ij)},S_{\rm tr}^{\E (ij)}$) and mixed ($S_{\rm int}^{X (ij)},S_{\rm tr}^{X (ij)}$) noise as a function of the dc offset of the driving. Inset in panel b) Line-shape of the applied voltages: harmonic driving, $V^{\rm h}_{\rm L}(t)=\bar{V}+V_{0}\cos(\Omega t)$, and bi-harmonic driving of the form $V^{\rm bh}_{\rm L}(t)=\bar{V}+V_{0}\cos(\Omega t)+\frac{V_{0}}2\cos(2\Omega t)$.
In all panels $k_{\rm B}T=0$ and $eV_{0}=2\hbar\Omega$. Full lines correspond to the case of harmonic driving while dashed lines represent the case of bi-harmonic driving.}
\label{fig}
\end{figure}

 The interference part of the energy noise, $S^{\E}_{\rm int}=\sum_{ij}S_{\rm int }^{\E(ij)}$ reads
\begin{equation}\label{eq:absorb}
S^{\E}_{\rm int}=\frac{D^{2}}{h}\left[\frac{2\pi^{2}\left(k_{\rm B}T\right)^3}{3}+\sum_{k=-\infty}^{+\infty}\frac{\left|e\, v_{{\rm L}k}\right|^{2}}{2}k\hbar\Omega\coth\left(\frac{k\hbar\Omega}{2k_{\rm B}T}\right)\right],\nonumber
\end{equation}
with  $v_{{\rm L}k}=\int_{0}^{\T}\frac{dt}\T V_{\rm L}(t)e^{ik\Omega t}$.  While the first term is again only given by thermal fluctuations, the second one is given by all the inelastic processes due to the ac driving field.
 This second term corresponds to the variance, $|e v_{{\rm L}k}|^2/2$, of the energy of a classical, charged particle in the oscillating potential $v_{{\rm L}k}\cos(k\Omega t)$ (i.e. the $k$th Fourier component of $V_\mathrm{ac}(t)$),  multiplied by the characteristic rate at which electrons exchange the energy $k\hbar \Omega $ with the ac field by fluctuating between states in the same reservoir~\cite{Battista14}.
In Fig.~\ref{fig}c) we plot the interference contributions $S^{\E (ij)}_{\rm int}$ to the energy noise as a function of the dc offset of the driving. The correlators between different kinds of particles, $S^{\E \rm (eh)}_{\rm int}=S^{\E \rm (he)}_{\rm int}$, have the same order of magnitude as the ones for equal kinds of particles, as long as the dc component of the driving signal is very small, but they decay rapidly for increasing $\bar{V}$. Depending on the sign of $\bar{V}$ and on the shape of $V_{\rm ac}(t)$ we find $S_{\rm int}^{\E \mathrm{(ee)}}> S_{\rm int}^{\E \mathrm{(hh)}}$ or $S_{\rm int}^{\E \mathrm{(ee)}}<S_{\rm int}^{\E \mathrm{(hh)}}$. 

The shape of the ac driving also determines the features of the transport contribution to the energy noise,  $S^{\E}_{\rm tr}=\sum_{i}S^{\E(ii)}_{\mathrm{tr}}$,  given by
\begin{equation}
S^{\mathcal{E}}_{\mathrm{tr}}= \frac{D(1-D)}{3h} \sum_{\ell=-\infty}^{+\infty}|c_{\rm L \ell}|^{2}\coth\left(\frac{\ell\hbar\Omega+e\bar{V}}{2k_{\rm B}T}\right)\left[(\ell\hbar\Omega+e\bar{V})^3+(\ell\hbar\Omega+e\bar{V})(\pi k_{\rm B}T)^2\right].
\end{equation} 
This quantity depends not only on the number of electron and hole excitations but also on their  energy  distribution, determined by $V(t)$: 
for example, if  $\max{V_{\rm ac}(t)}> |\min{V_{\rm ac}}(t)|$, the spread in energy of the electron-like excitations is larger than the one of the corresponding holes and this leads to $S_{\rm tr}^{\E{\rm (ee)}}>S_{\rm tr}^{\E{\rm (hh)}}$, as shown in the case of biharmonic driving in Fig.~\ref{fig}d). 
Differently  from what observed for the charge noise, the suppression of the creation of charge-neutral excitations (electron-hole pairs) is therefore not enough to  lead to a minimum of $S_{\rm tr}^{\E}$. It has been shown in Ref.~\cite{Battista14} that a trade-off between the number of the excitations and the energy range they span has to be reached in order to minimize the transport part of the energy noise, see Fig.~\ref{fig}d). 


\section{Mixed noise}\label{sec:mixedn}
We now turn to the  mixed correlator~\cite{Crepieux14}.
The total interference contribution, $S^{X}_{\rm int}=\sum_{ij}S_{\rm int }^{X (ij)}$ is
\begin{eqnarray}\label{eq:SintX_tot}
S^X_{\rm int}&=&\frac{e}{h}D^{2}e\bar{V}k_{\rm B}T.
\end{eqnarray}
The interference part of the mixed noise shows different features from $S^{}_{\rm int}$ and $S^{\E}_{\rm int}$.  It is proportional to the interference part of the charge noise $S_{\rm int}$ times $e\bar{V}/2$, which is the average energy of the electrons  injected into the system at the left contact, with respect to $V_\mathrm{R}=\mu=0$. It is hence non-vanishing only if  the system is driven out of equilibrium and  temperature is finite.
\indent In Fig.~\ref{fig}e) we plot the contributions to $S^X_{\rm int}$ from excitations of the same or of different kind, $S^{X (ij)}_{\rm int}$, as a function of $e\bar{V}$ at zero temperature. We find $S^{X (\rm ee)}_{\rm int}=-S^{X (\rm he)}_{\rm int}$ and $S^{X (\rm hh)}_{\rm int}=-S^{X (\rm eh)}_{\rm int}$ giving as a consequence that the interference contribution to the correlator between fluctuations of the energy carried by the electrons (holes) and the total charge current is zero. \red{Moreover, we have in general $S^{X (\rm ee)}_{\rm int}\neq S^{X (\rm hh)}_{\rm int}$, apart for the case when $e\bar{V}=0$ and $V_{\rm ac}(t)$ is symmetric with respect to one of its nodes, so that the holes and the electrons are symmetrically distributed in energy with respect to $\mu=0$. In that case they just have opposite sign. If an asymmetry in the energy distribution of the electron- and hole-like excitations is introduced {\em only} as an effect of the dc-bias,  see e.g. the case of the harmonic signal shown in Fig.~\ref{fig}e),  then we have $S^{X (\rm ee)}_{\rm int}(e\bar{V})=-S^{X (\rm hh)}_{\rm int}(-e\bar{V})$. }

The transport contribution, $S^{X}_{\rm tr}=\sum_{ij}S^{X (ii)}_{\rm tr}$, is given by
\begin{equation} 
S^{X}_{\mathrm{tr}}= \frac{eD(1-D)}{2h} \sum_{\ell=-\infty}^{+\infty}|c_{\rm L \ell}|^{2} (\ell\hbar\Omega+e\bar{V})^2\coth\left(\frac{\ell\hbar\Omega+e\bar{V}}{2k_{\rm B}T}\right)\label{eq:SEtrT}
\end{equation}
At zero temperature, it is the only non-vanishing contribution to the mixed noise, $S^X(k_{\rm B}T=0)=S^X_{\mathrm{tr}}(k_{\rm B}T=0)$. We can separate a contribution due to electron-like excitations from the  one due to hole-like excitations 
\begin{equation}\label{eq:SiiT0X}
S^X(k_{\rm B}T=0)=\sum_{i=\rm e,h}\frac{\Omega}{2\pi}D(1-D)\Xi_{i}.
\end{equation}
The quantity $\Xi_{i}$ is formally similar to the definition of $N_i$,
\begin{equation}
\Xi_{i}=\sigma_{i}e\sum_{\ell=-\infty}^{+\infty}|c_{\rm L \ell}|^{2}\frac{(\ell\hbar\Omega+e\bar{V})(\ell+e\bar{V}/\hbar\Omega)}{2}\theta_{i}(\ell \hbar\Omega+e\bar{V}),
\end{equation}
 but each contribution coming from a different subband $l$ is weighted by its corresponding energy. It can thus be understood as the total average energy transported by $i$-like excitations multiplied by $\sigma_{\rm e/h}e=\pm e$ the corresponding charge. Indeed, $(\Xi_{\rm e}-\Xi_{\rm h})/(e\mathcal{T})=\overline{eV_\mathrm{L}(t)^2}/2h$ corresponds to the  power $P$  dissipated to one of the leads per period.
In a two-terminal conductor characterized by energy-independent transmission probability at the central scatterer, the energy is equally dissipated in the two reservoirs resulting in the factor $1/2$. 

In Fig.~\ref{fig}f) we plot $S^{X(ii)}_{\rm tr }$ at zero temperature as a function of the dc part of the driving signal.
\red{The correlators  $S^{X(\rm ee)}_{\rm tr }$  and $S^{X(\rm hh)}_{\rm tr }$ have opposite signs, as a consequence of the fact that electron- and hole- excitations have different charges but both give a contribution to the energy current of the same sign. Moreover, they become negligible when 
$e\bar{V}\ll0$  or $e\bar{V}\gg0$, respectively, since  $\Xi_{\rm e}$ and $\Xi_{\rm h}$ are suppressed in these voltage ranges. This leads to a change of sign of the total transport contribution $S^{X}_{\rm tr }$ as a function of  $e\bar{V}$, in contrast to $S_{\rm tr }$ and $S^{\mathcal{E}}_{\rm tr }$.}

We now relate the mixed noise to  the zero frequency correlator between the power fluctuations and the charge current fluctuations, $\overline{\langle \Delta P(t)\Delta \hat{I}_{\rm R}(t+\tau)\rangle}=\int_{0}^{\mathcal{T}}dt\int_{-\infty}^{\infty}d\tau\langle\{\Delta \hat{P}(t),\Delta \hat{I}_{\rm R}(t+\tau) \}/(2\mathcal{T})$, a measurable quantity. Here, we define $\hat{P}(t)=-V_\mathrm{L}(t)\hat{I}_\mathrm{L}(t)$, the operator for the power provided by the time-dependent voltage source. 
We find $\langle \Delta P(t)\Delta \hat{I}_{\rm L}(t+\tau)\rangle+\langle \Delta P(t)\Delta \hat{I}_{\rm R}~(t+~\tau)~\rangle=0$, since neither charge nor energy can be accumulated in the conductor. Explicitly, we have 
\begin{equation}\label{eq:potcurrcorr}
\overline{\langle \Delta P(t)\Delta I_{\rm R}(t+\tau)\rangle}=
2S^X_{\rm int}+2S^X_{\rm tr},
\end{equation}
with the mixed correlator contributions discussed in Eqs.~(\ref{eq:SintX_tot}) and (\ref{eq:SEtrT}).
Importantly,  for a completely transparent scatterer, $D=1$, $2S^X_{\rm int}$ is the only contribution to the zero-frequency correlator of power and charge current fluctuations; if the transparency of the scatterer is reduced, the second term which corresponds to $ 2S^{X}_{\rm tr}$, is the dominant one.   Both terms should  thus be well observable.
\section{Conclusions}\label{sec:conclu}
We considered a two-terminal conductor subjected to a time-periodic bias voltage. We discussed the correlator of charge current fluctuations and energy current fluctuations and the correlator between charge and energy current fluctuations.  All the considered correlators show a transport contribution, due to correlations between particles exchanged between the two leads, and an interference contribution, due to the exchange of particles between different energy states in the same lead. The mixed noise is nonzero since electrons and holes are carriers both of charge and energy.  At zero temperature, it is formally analogous to the charge noise, but now what is partitioned at the scatterer is not the number of impinging electrons or holes, but their energy, averaged over the different subbands.   In contrast with charge and energy noise, it does not show a minimum as a function of an applied dc bias,  but it changes sign when the bias is reversed. All discussed effects are expected to be observable by looking at the correlations  between power  and charge current fluctuations, under given conditions.


%

\end{document}